\documentclass{webofc}
\usepackage[varg]{txfonts}   % Web of Conferences font
\usepackage{hyperref}
\usepackage{url}

\hypersetup{colorlinks=true,citecolor=blue,urlcolor=blue,linkcolor=blue}
\begin{document}
\title{Improvements of the GPU Processing Framework for \hbox{ALICE}}

\author{\firstname{David} \lastname{Rohr}\inst{1}\fnsep\thanks{\email{drohr@cern.ch}}
        % etc.
}

\institute{CERN,\\
  1211 Geneva,\\
  Switzerland
          }

\abstract{
ALICE is the dedicated heavy ion experiment at the LHC at CERN and records lead-lead collisions at a rate of up to 50 kHz.
The detector with the highest data rate of up to 3.4 TB/s is the TPC.
ALICE performs the full online TPC processing corresponding to more than 95\% of the total workload on GPUs, and when there is no beam in the LHC, the online computing farm's GPUs are used to speed up the offline processing.
After the deployment of the first version of the online TPC processing needed for data taking, ALICE has implemented many improvements to its GPU processing framework.
These include a run time compilation mode applying on the fly optimizations, improvements to parallelize / speed up the GPU compilation, debugging modes to guarantee reproducible and deterministic results in concurrent reconstruction, and framework features to leverage common components in the code of different detectors.
The talk will give an overview of the ALICE experience with GPUs in online and offline processing and present the latest GPU processing framework features.
}
\maketitle
\section{Introduction}

ALICE (A Large Ion Collider Experiment) \cite{bib:alice} at CERN in Geneva is the dedicated heavy ion experiment at the LHC (Large Hadron Collider).
During the Long Shutdown 2 (LS2), it was upgraded~\cite{bib:ls2upgrade} to run in a continuous readout mode and both hardware and software triggers were removed for most subdetectors.
Today in Run 3, ALICE records heavy ion collisions at up to the LHC's peak nominal interaction rate of 50 kHz storing all data.
This is two orders of magnitude more than in Run 2 before LS2.
This necessitates a very efficient data reduction scheme in the absence of a trigger, since the storage capacity cannot be increased by the same magnitude as the recorded interaction rate.
An update of the whole online computing scheme \cite{bib:o2tdr}  and infrastructure was performed in parallel to the detector upgrades during LS2, installing a computing farm consisting of 350 servers equipped with 8 GPUs each.
With the change from a triggered to a continuous readout, the basic processing unit is no longer an event but a time frame consisting of a configurable period of continuous data.
Currently, ALICE uses time frames of 32 LHC orbits lasting around 2.5\;$\mu$s.

During data taking, the online computing farm performs real time calibration, quality control, and most importantly data compression, while in time periods without beam in the LHC or during pp data taking when not all online nodes are needed for online processing, it operates as a GRID site performing offline reconstruction.
The detector with the largest data volume by far is the TPC (Time Projection Chamber), thus TPC processing and compression is the most critical part during online processing.
ALICE has developed the O$^{\text{2}}$ framework as unified software framework for online and offline reconstruction, simulation, and analysis, and has adopted GPU reconstruction to speed up the critical parts of event reconstruction \cite{bib:chep2023}.
The driving factor for using GPUs for the parallelization is the fact that TPC online processing, which is the most computing-intense task during data taking, is highly parallelizable and well-suited for GPU offload.
In order to be independent from vendors and GPU programming languages, ALICE is using a custom abstraction layer on top, with the vast majority of the algorithms being implemented in pure generic \texttt{C++} \cite{bib:generic}.

For the best use of the computing resources of the online farm during offline processing, as the first LHC experiment ALICE has started to use GPUs for offline processing beginning with the reconstruction of 2022 data.
ALICE has demonstrated a GPU speedup of 2x for Pb--Pb data and 2.5x for pp data on its online farm \cite{bib:ichep2024}.
In contrast to online processing, which is fully bound by the TPC GPU processing, offline reconstruction requires much more CPU time and is currently CPU-bound.
GPU idle times are currently above 50\%.
ALICE aims to offload more offline processing steps to GPUs in the future.
The aim is to achieve a 5x GPU speedup for both online and offline by offloading the full barrel tracking to GPU \cite{bib:chep2023,bib:ichep2024}.

In order to use the GPUs in a broader scope, for a larger set of algorithms and on more different hardware platforms on different GRID sites in the future, ALICE is working on general improvements to its GPU framework.
Some of them will be presented in the following sections.

\section{Deterministic processing mode for validation and debugging}

For development, debugging, and validation of the GPU code it is extremely helpful if results are reproducible and can be compared one-to-one on a bit level to the host.
In previous work~\cite{bib:chep2012}, we concluded that there will be some indeterminism in the code simply from using mathematics optimization from compiler options, like the assumption that floating point arithmetic is associative.
This led to differences e.\,g.~in cluster to track association affecting far less than $0.001$\% of the clusters with absolutely no impact on the physics results.
Consequently, also other small indeterministic effects, e.\,g.~coming from concurrent processing, were accepted in ALICE code.

This approach has not changed for high-performance code in production.
However, so far this made results indeterministic and prevented binary comparisons of the output.
Thus, ALICE developed a special optional deterministic mode for its GPU code, which guarantees fully deterministic results and enables comparing GPU to CPU results bitwise one-to-one.
This comes at the cost of partially greatly increased processing time with a factor between 1.5 and 10.
It is important to note that the performance of the default mode is not changed at all.

The deterministic mode is implemented as a combination of compile time and run time changes, i.\,e., it needs a special compilation of the ALICE software and a switch at run time.
The goal is to implement the run time changes as additional kernels (e.\,g.~for sorting) executed after normal processing kernels to minimize the changes to the normal processing.
By design, this is not possible for the compile time changes, e.\,g.~different compiler options.
The possibility that the deterministic mode hides some problems in this way is unavoidable, but it affects only few rare cases.
The deterministic mode has been highly successful in debugging.
It uncovered several long-standing bugs in the ALICE code that were previously attributed to parallel processing.
On top of that, it revealed several compiler bugs, which were reported and subsequently fixed.

\noindent
\ \newline
The deterministic mode implies the following changes at compile time:

\begin{itemize}
 \item All sorting must apply a total ordering. Partial sorting operators are amended by arbitrary comparisons to guarantee a total order.
 \item Fused Multiply Add (FMA) is disabled and the \texttt{-ffast-math} compiler flag is removed, ensuring IEEE-compliant floating point processing.
 \item All other fast mathematics optimizations, like fast inverse square root, are disabled (by default there can be different architecture-dependent optimizations).
 \item Denormalized floats are disabled and set to zero in a consistent way.
 \item All floating point functions that do not have a well defined deterministic output, such as trigonometric functions, are made deterministic. Since the GPU code uses only single-precision floating point, it is sufficient to replace such functions with their double-precision counterparts and then round back to single precision.
 \item In some rare cases, algorithms that are intrinsically indeterministic are replaced by deterministic ones at compile time.
\end{itemize}

 \noindent
The following settings can be enabled via a switch at run time (i.\,e.~they are always available, even for normal software builds):

\begin{itemize}
 \item Additional sorting kernels are executed on the GPU after each kernel that has outputs whose order depends on the concurrency. The sorting guarantees a deterministic order, maintaining all relations like cluster to track association indices.
 \item In some cases parallel processing is disabled and the processing runs single-threaded.
 \item In some cases, slower deterministic algorithms are executed instead of optimized indeterministic ones.
\end{itemize}

\section{Per-kernel compilation units}

\label{perkernel}

So far, ALICE had aggregated all GPU kernels in a single file automatically during the CMake build process, and then compiled this as a single compilation unit.
This proved to be efficient at the beginning, but has the drawback that it creates a single long-lasting serial compilation step during the software build, which grows with every GPU kernel added to O$^{\text{2}}$.
Since this started to become a bottleneck with more than 100 GPU kernels during O$^{\text{2}}$ compilation today, an effort started to switch to multiple compilation units.

In a first attempt, CUDA\footnote{The NVIDIA CUDA GPU framework: https://developer.nvidia.com/cuda-toolkit} and ROCm\footnote{The AMD ROCm Software: https://www.amd.com/de/products/software/rocm.html} RDC (Relocatable Device Code) was used, which allows to compile GPU source files as individual compilation units.
The object files are then linked together in a device linking step, in the same way as compilation and linking is split for \texttt{C++} host code.
Unfortunately, this led to several issues. Most importantly, GPU code is compiled for a compile time defined number of maximum registers, which can differ per kernel.
With RDC, device functions in a compilation unit are compiled only once with a fixed number of allowed registers, and it is then used by all kernels with this number.
More importantly, at compile time the compiler does not know which kernels the function will be used in, so it must guess the number of registers, which eventually might exceed the maximum number available in a kernel, which will lead to linking or run time failures.

This can be overcome by manually defining the number of registers for such functions.
But besides this being a tedious manual process, it is not easy for the developer to learn what is a good number, and more importantly, the optimal number can actually be different for different kernels.
Some first tests showed on the order of 10\% performance degradation for the ALICE code in this mode.
Thus, besides the fact that RDC compilation with manual register specification works for ALICE and is still available if desired by developers, a different approach was followed.

Via CMake scripts, ALICE O$^{\text{2}}$  prepares one GPU source file per kernel, which contains all the includes for that kernel, and these files are compiled as individual compilation units.
This yields optimal register numbers for each functions, and even produces slightly faster code than having all kernels in one compilation unit.
Technically, CMake calls the NVIDIA CUDA compiler \texttt{nvcc} and the AMD ROCm compiler \texttt{hipcc} multiple times directly, once per kernel, to directly create GPU binary code.
The GPU binaries are then loaded via \texttt{cuModuleLoad} and \texttt{hipModuleLoad}.
Since many functions are now compiled multiple times in different compilation units, the cpu time increases to some extent, but the wall time is decreased significantly by the parallel compilation.

A related problem that appeared with multiple compilation units in this way is that in CUDA constant memory symbols in non-RDC code are defined once per compilation unit.
Thus, the new approach multiplies the constant memory buffers by the number of compilation units, currently around 100.
The approach to deal with this is described in the next section.

\section{Sharing components between GPU libraries}

\label{sharedcomponents}

The ALICE GPU framework provides some generic components, which are needed by several detectors and algorithms.
For instance, TPC tracking, ITS tracking, and vertex fitting all require track propagation taking into account the magnetic field map and the material budget map.
For a better organization of the code, ALICE has started to split the GPU code to different libraries in different Linux shared object files.
A question is what is the best approach, to provide such shared components to all libraries.
The canonical way is to use RDC, and provide an object file that can be linked later.
This is the only way in which one can avoid multiple compilation of the files.
Obviously, it comes with the caveat described above, that the register usage will not be optimal in all kernels.

\ \newline
\noindent
ALICE now provides two options:
\begin{itemize}
\item To guarantee maximum performance, libraries can integrate all the code into their kernels.
This is what is done for the TPC code, which is most performance-critical.
\item Alternatively, a GPU RDC library is provided as CMake object library, to which the shared libraries can link which will automatically provide all the requirements.
\end{itemize}

\subsection{Initializing constant memory buffers and keeping them in sync}

\label{constmem}

A problem that comes with the approaches of sections \ref{perkernel} and \ref{sharedcomponents} is that GPU code is compiled multiple times.
If the code has constant memory buffers, this creates one instance of the buffer per compilation unit.
The buffers of the global components contain constants as for TPC geometry, and calibration constants.
The content is always constant for a time frame but may change between time frames.

To make this transparent to the user, the global GPU framework keeps track of the constant memory buffers and keeps them up to date.
In particular, using weak symbols and static instances, the framework code in the CMake object library will automatically register all constant memory symbols to the global GPU framework.
For using shared components, the user does not need to do anything beyond linking to the CMake object library.

The GPU framework will initialize all constant buffers at the beginning, and will update all of them when needed.
A downside of this approach is that with more than 100 kernels, the same buffer needs to be updated more than 100 times instead of once, since all of the kernels will use the same constant.
Unfortunately, there is currently no better way, since CUDA by design has one constant buffer per compilation unit.
However, the performance impact of this implementation has been evaluated to be completely negligible.
In the end, these are buffers in the order of 100 64\,KB, which are updated once per time frame, with a processing time of the order of 1\,s per 32 orbit time frame on e.\,g.~an NVIDIA 3090 GPU.

\section{On the fly optimization with Run Time Compilation (RTC)}

\label{RTC}

There are plenty of variables, which will be constant during run time, once the processing task are initialized.
In particular, these settings will control which code branches are taken in many places.
Fixing these settings at compile time allows to eliminate a lot of branching.
In particular GPUs are sensitive to branches and complex code, even if the run time will always follow the same branch, since GPUs do not have sophisticated branch prediction.

Therefore, ALICE has implemented a framework to optimize the code on the fly when all the run-time-constant settings are known.
Using RTC (Run Time Compilation), binary code is produced that eliminates much unnecessary branching.
For this, the GPU processing library contains the source code of its kernels.
During initialization, it replaces many variables in the code by \texttt{C++} constexpr constants, and recompiles the code, such that the compiler performs dead code elimination to reduce the branching.
For the compilation, loading of the GPU binary code, and keeping the constant memory buffers up to date, the same infrastructure is used as described in section \ref{perkernel} for module loading and in section \ref{constmem} for constant memory.

On top of that, there is the possibility to use preprocessor macros in the code, which can enable, disable, or modify code during RTC.
In particular, this is used to force some calibration constants to 0 during online processing, which are only needed for offline reconstruction, and which created a large performance impact for the online TPC processing when introduced, even if the code is not executed.
This is a prime example of complex code and branching making GPU code slower, even if it takes always the same branch, while the impact on CPU performance is only on the order of 1\%.
For details see Fig.~\ref{perftable} in section~\ref{evolution}.

\subsection{Using RTC on the online computing farm}

Some special handling is done for the online computing farm.
ALICE uses 8 GPUs driven by 8 operating system processes~\cite{bib:chep2023}, which all need the RTC code.
Instead of compiling multiple times, file locks and a Linux temporary in-memory file system (\texttt{tmpfs}) are used to compile only once and then distribute the code to all processes.
Since the software is running in job containers, it must be ensured that the GPU compiler is available to the job and certain environment variables e.\,g.~for temporary files must be set correctly.
And finally, the processes driving the GPUs are pinned to NUMA (Non Uniform Memory Architecture) domains.
In order to use all CPU cores for GPU RTC compilation, the pinning must be released for the compilation jobs spawned by the GPU process.

\subsection{Architecture support with CVMFS builds}

Compilation of the GPU code for a particular architecture prolongs the software build by 3 to 5 minutes per GPU architecture.
Therefore, currently the ALICE software builds that are published to CVMFS contain only binary code for 3 architectures, 2 for AMD and 1 for NVIDIA.
More architectures are not compiled in to keep the compilation time short.
The support can be widened by adding code for virtual architectures, which is not the final binary code and is then compiled for the exact hardware at run time by the GPU driver.
However, this introduces a minor performance penalty, since the compiler cannot fully optimize for the target architecture.
Instead, ALICE has the option to use its RTC feature to compile at run time for the local architecture of the server the software is running on.
Tests are currently ongoing to support GRID sites with different GPU architecture in this way.

\section{Evolution of the GPU processing speed in 2024}

\label{evolution}

\begin{figure*}[b]
\centering
\vspace*{1cm}
\includegraphics[width=13cm,clip]{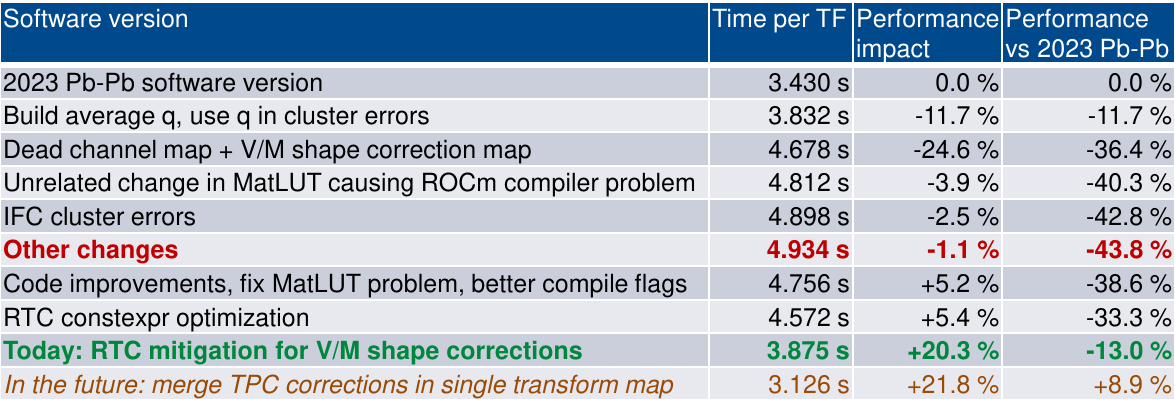}
\caption{
Evolution of ALICE TPC GPU online processing time per time frame:
the relative speedups / degradations in columns 3 and 4 are normalized to the performance during 2023 Pb--Pb data taking (first line), with the performance impact (column 3) showing the performance of the feature in a line compared to the previous line, while the Performance v.\,s.~2023 Pb--Pb (column 4) shows the relative impact of the software including this feature and all previous features v.\,s.~the software version running in November 2023.$^{\text{3}}$
}
\label{perftable}
\end{figure*}

Fig.~\ref{perftable} gives an overview of the evolution of the processing speed in 2024.\footnote{
Note that Fig.~\ref{perftable} sums up effects of individual commits related to a topic, not following the real chronological order of commits, thus ignoring the interplay of the commits.
Therefore the numbers are not accurately representing the time evolution but are meant to give an overview of the impact of different features.
Measuring step by step would give other percentages.
The measurements were performed on a reference MC time frame with 100 Pb--Pb collisions on an online computing node with an AMD MI50 GPU.
\label{perftablenote}}
After the Pb--Pb data taking at the end of 2023, several improvements were added to the TPC reconstruction code for better reconstruction efficiency.
This includes in particular better cluster error estimation and the usage of a dead channel map.
In addition, the TPC cluster coordinate transformation was amended by several corrections that are only used offline.
Unfortunately, these corrections had a negative performance impact on the online processing on the order of 20\% simply by making the TPC code more complicated.
Using RTC (see section RTC), this performance degradation could be fully mitigated for the Pb--Pb data taking 2024.
Overall, the 2024 Pb--Pb processing performance was a bit slower than 2023, but in the margin of the online computing farm capacity.
Currently, ALICE is developing a general improvement of the TPC correction code, which will not need this mitigation any more, and which has already been demonstrated to yield ~10\% faster execution for 2025.

\section{Conclusions}

For Run 3 ALICE has updated its detectors, its online computing farm, and its online and offline processing software significantly.
ALICE was the first LHC experiments to use GPUs for online processing, and does so heavily in Run 3 within the O$^{\text{2}}$ software framework.
As first LHC experiment, ALICE is also using GPUs for GRID jobs starting with the processing of 2022 data.
In the online processing, there is 99\% of the processing load on GPUs, being fully GPU bound, with smooth operation from 2022 to 2024.
Today, 50\% to 60\% of the offline processing is offloaded to GPUs achieving a 2x to 2.5x speedup in Pb--Pb and pp.
ALICE aims to have the full barrel tracking on GPUs eventually corresponding to 80\% of the total workload with an estimated speedup of 5x.

In 2024, ALICE has developed several new features for the GPU processing.
These features will in particular simplify the adaptation and porting of more reconstruction steps to GPU and they simplify the debugging and validation.
Most notably, the per-kernel compilation speeds up the compile time, the deterministic mode simplifies debugging, validation, and development, framework support for using shared components in external libraries simplifies the development, and run time compilation provides a global increase of processing speed.

For 2025 ALICE plans to extend the GPU usage in the GRID to more architectures, plans to port more processing steps to GPU, and aims for significant speed-ups of some of its GPU code.

\end{document}